\documentstyle[a4,12pt,epsf,amsfonts,amssymb]{article}  

\DeclareFontFamily{U}{rsfs}{\skewchar\font"7F} 
\DeclareFontShape{U}{rsfs}{m}{n}{<-6> rsfs5 <6-8> rsfs7 <8-> rsfs10}{} 
\DeclareMathAlphabet{\mathscr}{U}{rsfs}{m}{n}  

\def\ben{\begin{equation}} 
\def\een{\end{equation}}
\def\bena{\begin{eqnarray}} 
\def\eena{\end{eqnarray}}
\def\f(#1/#2){\frac{#1}{#2}} 
\def\Frac(#1/#2){\left(\frac{#1}{#2}\right)}
\def\scalar{{\Bbb S}}

\def\Dphi{{\mit \Delta}\phi}

\def\half{{1\over 2}}

\def\Tdot#1{{{#1}^{\hbox{.}}}}

\def\dddot#1{\stackrel{...}{#1}{}\!\!}

\newcommand{\la}{\langle}
\newcommand{\ra}{\rangle}

\newcommand{\D}{{\mathcal D}}

\newcommand{\non}{\nonumber}    

\begin{document} 

\title{ Can the Acceleration of Our Universe 
Be Explained by the Effects of Inhomogeneities? }  

\author{Akihiro Ishibashi$^{\dag}$ and Robert M. Wald$^{\dag \ddag}$ \\ \\  
 {\it Enrico Fermi Institute$^{\dag}$ and Department of Physics$^{\ddag}$}\\  
 {\it The University of Chicago, Chicago, IL 60637, USA} 
        } 


\maketitle

\begin{abstract}
No. It is simply not plausible that cosmic acceleration could
arise within the context of general relativity from a back-reaction
effect of inhomogeneities in our universe, without the presence of a
cosmological constant or ``dark energy.''  We point out that our 
universe appears to be described very accurately on all scales by a
Newtonianly perturbed FLRW metric. (This assertion is entirely
consistent with the fact that we commonly encounter $\delta \rho/\rho
> 10^{30}$.) If the universe is accurately described by a Newtonianly
perturbed FLRW metric, then the back-reaction of inhomogeneities on the
dynamics of the universe is negligible. If not, then it is the burden
of an alternative model to account for the observed properties of our
universe. We emphasize with concrete examples that it is {\it not}
adequate to attempt to justify a model by merely showing that some
spatially averaged quantities behave the same way as in FLRW models
with acceleration. A quantity representing the ``scale factor'' may
``accelerate'' without there being any physically observable
consequences of this acceleration.  It also is {\it not} adequate to
calculate the second-order stress energy tensor and show that it has a
form similar to that of a cosmological constant of the appropriate
magnitude. The second-order stress energy tensor is gauge dependent,
and if it were large, contributions of higher perturbative order could
not be neglected. We attempt to clear up the apparent confusion
between the second-order stress energy tensor arising in perturbation
theory and the ``effective stress energy tensor'' arising in the
``shortwave approximation.''
\end{abstract} 

\section{Introduction} 

The apparent acceleration of our universe is one of the most striking
cosmological observations of recent times. In the context of
Friedmann-Lemaitre-Robertson-Walker (FLRW) models in general
relativity, the acceleration of our universe would require either the
presence of a cosmological constant or a new form of matter (``dark
energy'') with large negative pressure. However, since there does not
appear to be any natural explanation for the presence of a
cosmological constant of the necessary size, nor does there appear to
be any natural candidate for dark energy, it is tempting to look for
alternative explanations.  In recent years, there have been at least
two approaches that have attempted to account for the observed
acceleration of our universe within the framework of general
relativity as being a consequence of deviations of our universe from
exact FLRW symmetry, without invoking the presence of a cosmological
constant or dark energy.

One approach notes that the mass density of our universe is, in fact,
extremely inhomogeneous on scales much smaller than the Hubble radius.
In order to get an effective homogeneous, isotropic universe, one needs
to average and/or smooth out the inhomogeneities on some appropriate
choice of spatial slicing. In such an ``averaged'' (or effective)
FLRW universe, one can then define 
``effective cosmological parameters''~\cite{Buchert00,Buchert01,BC03}. 
One then finds that the equations of motion for these effective cosmological 
parameters differ, in general, from the equations satisfied by these 
parameters in FLRW models. 
If they differ in a way that corresponds to adding a cosmological 
constant (or dark energy) of the right magnitude, then one may 
hope to have explained the acceleration of our
universe in the context of general relativity, without invoking the presence 
of a cosmological constant or dark 
energy~\cite{Rasanen03,KMNR05,KMR05,BMR05,Nambu05a,NT05,Moffat05}.

A second approach has attempted to account for the acceleration 
of our universe as a back-reaction effect of long-wavelength cosmological 
perturbations. Here one constructs an ``effective energy-momentum tensor'' 
for these perturbations from second-order perturbation theory, 
and adds this as a source term in Einstein's 
equation~\cite{MAB1997Lett,ABM1997PRD,Nambu02,BL2004}. 
If this effective energy-momentum tensor has a form similar to that 
of a cosmological constant (or, at least, provides a negative 
pressure term) and is of the appropriate magnitude, then one may hope that 
this could explain the acceleration of our universe.

Given that our universe appears to be very accurately described by a
FLRW model, with very small deviations from homogeneity and isotropy,
it would seem extremely implausible that cosmologically important
effects could result from the second (or higher) order corrections
produced by these small departures from a FLRW model.  The main
purposes of this paper are to explain this point with somewhat more
precision and to point out some significant flaws in the arguments
that have been made in the context of the above two approaches.  In
particular, we emphasize that one cannot justify a model by merely
showing that spatially averaged quantities behave the same way in FLRW
models with acceleration; rather one must show that {\it all} of the
predictions of the model are compatible with observations. We
illustrate this point by showing that spatial averaging can yield
``acceleration'' for the case of a universe that consists of two
disconnected components, each of which is decelerating! We also show
that spatial averaging can yield acceleration for suitably chosen
slices of Minkowski spacetime. With regard to the back-reaction
effects of long-wavelength perturbations, we note that the effective
energy-momentum tensor is highly gauge dependent, as has previously
been pointed out by Unruh~\cite{Unruh1998}. Even in the long-wavelength 
limit, we show that one can get essentially any answer one wishes 
for the effective energy-momentum tensor, even though
one cannot get any new physical phenomena beyond those
already present in FLRW models. We also emphasize the difference
between the use of an effective energy-momentum tensor for
gravitational perturbations in the context of second-order
perturbation theory and the use of a similar effective energy-momentum
tensor in the context of the ``shortwave
approximation''\cite{BH1964,Isaacson1968}. 
The former is highly gauge dependent (and, thus, not easily interpreted) 
and must be ``small'' if higher order perturbative corrections are to be 
neglected. The latter is essentially gauge independent and need not be
``small.'' However, the ``shortwave approximation'' clearly is not valid for
analyzing long-wavelength cosmological perturbations.

In the next section, we point out that our universe appears to be very
accurately described on all scales by a Newtonianly perturbed FLRW
metric, despite the presence of large density contrasts.  If this is
correct, then higher order corrections to this metric resulting from
inhomogeneities would be negligible. Thus, any model that attempts to
explain the acceleration of our universe as a consequence of higher
order effects of inhomogeneities will have to overcome the seemingly
impossible burden of explaining why the universe appears to be so well
described by a model that has only very small departures from a FLRW
metric.\footnote{ 
For example, rotation of the cosmic matter may produce acceleration 
effects~\cite{KMNR05} but these acceleration effects must be negligible 
in view of the observed isotropy of our 
universe~\cite{Flanagan05,HS05,GCA05}. 
} 
In section 3, we illustrate that spatial averaging can produce
an entirely spurious ``acceleration'' that is not associated with any
physical observations. In section 4, we 
emphasize the distinction between
second-order perturbation theory and the shortwave approximation, and
we analyze the gauge
dependence of the effective energy-momentum tensor arising in
second-order perturbation theory.

\section{The Newtonianly Perturbed FLRW Metric} 

By a Newtonianly perturbed FLRW metric, we mean a metric of the form
\ben 
  d\tilde{s}^2 = -(1 + 2 \Psi)dt^2 + a^2(t)(1- 2 \Psi)
  \gamma_{ij}dx^idx^j \,.
\label{metric:Newton-FLRW}
\een 
where $\gamma_{ij}$ denotes the metric of a space of constant
curvature (3-sphere, flat, or hyperboloid), and $\Psi$ satisfies
\bena 
 |\Psi| \ll 1 \,, \quad 
 \left| \frac{\partial \Psi}{\partial t} \right|^2   
    \ll \frac{1}{a^2} D^i\Psi D_i \Psi \,, \quad 
 (D^i \Psi D_i \Psi)^2 \ll (D^iD^j\Psi) D_iD_j\Psi \,,   
\label{condi:Newtonian}  
\eena 
where $D_i$ denotes the derivative operator associated with
$\gamma_{ij}$. Suppose that the stress-energy content of this spacetime consists of fluid components
that are very nearly homogeneously and isotropically distributed
but may have arbitrary equation of state (such as radiation, dark energy, and/or a
cosmological constant)
together with components that may be very inhomogeneously distributed on scales 
small compared with the Hubble radius but are nearly
pressureless and move with velocity much smaller than light relative
to the Hubble flow (such as ordinary matter and dark matter).
The stress-energy of the smoothly distributed components take the form
\ben 
T^{(s)}_{ab} \approx \rho^{(s)}(t) dt^2
+ P^{(s)}(t) a^2(t) \gamma_{ij}dx^idx^j  \,, 
\label{Ts}
\een
where $P^{(s)} = P^{(s)}(\rho^{(s)})$ is arbitrary, 
whereas the inhomogeneously distributed components 
have a stress-energy of the ``dust'' form 
\ben 
T^{(m)}_{ab} \approx \rho^{(m)}(t,x^i) dt^2 \, .
\label{Tm}
\een

If one plugs the metric form eq.~(\ref{metric:Newton-FLRW})
into Einstein's equation and uses eqs.~(\ref{condi:Newtonian}),
the spatial average yields the usual FLRW equations for the scale 
factor $a$ with stress-energy source consisting of the sum of eq.~(\ref{Ts}) 
and the spatial average of eq.~(\ref{Tm}), 
\bena 
  3 \left(\frac{\dot a}{a}\right)^2 
  &=& \kappa^2 \left( \rho^{(s)} + \bar{\rho}^{(m)} \right) 
   - 3 \frac{K}{a^2} \,, 
\label{Einstein:tt}
\\
  3\frac{\ddot a}{a} 
 &=& -\frac{\kappa^2}{2}\left( 
                             \rho^{(s)} + \bar{\rho}^{(m)} + 3 P^{(s)} 
                        \right) \,,   
\label{Einstein:ij} 
\eena 
where $\bar{\rho}$ denotes the spatial average of $\rho$ (taken 
on a $t = {\rm const.}$ time slice with respect 
to the underlying FLRW metric with $K=\pm 1,\, 0$). 
The dominant remaining terms in Einstein's equation 
then yield 
\bena
  \frac{1}{a^2} \Delta_{(3)} \Psi = \frac{\kappa^2}{2} \delta \rho \,, 
\label{eq:Poisson}
\eena 
where $\Delta_{(3)} \equiv \gamma^{ij}D_i D_j$, and 
where $\delta \rho = \rho^{(m)} - \bar{\rho}^{(m)} $ denotes 
the deviation of the density from the spatial average. 
In the following we assume the metric $\gamma_{ij}$ to be a flat spatial 
metric, i.e., $K=0$.

It should be emphasized that the above discussion does {\it not} 
constitute a {\it derivation} that a metric form,  
eq.~(\ref{metric:Newton-FLRW}), is a good approximation 
to a solution to Einstein's equation for stress-energy eq.~(\ref{Ts})
and (\ref{Tm}) when eqs~(\ref{Einstein:tt})--(\ref{eq:Poisson}) 
hold. Rather, all that 
has been shown is that if one {\it assumes} a metric of the form
eq.~(\ref{metric:Newton-FLRW}) with eqs.~(\ref{condi:Newtonian})
holding and with the stress-energy given by eqs.~(\ref{Ts}) and
(\ref{Tm}), then eqs.~(\ref{Einstein:tt}), (\ref{Einstein:ij}) 
and (\ref{eq:Poisson}) must hold. 
The above discussion is thus analogous to the usual 
textbook ``derivations'' of the ordinary Newtonian limit 
of general relativity, where one also postulates a spacetime metric 
of a suitable ``Newtonian form'' and 
assumes that the matter distribution is approximately of the form
(\ref{Tm}) (see, e.g., section 4.4a of \cite{w84}). 
One can
derive the ordinary Newtonian limit more systematically by considering
one-parameter families of solutions to Einstein's equation with
suitable limiting properties 
(see, e.g., \cite{FS83} and \cite{ehlers} and references cited therein). 
It would be more difficult to
provide an analogous analysis here, but we see no reason to doubt that
eq.~(\ref{metric:Newton-FLRW}) is a good approximation to a solution
to Einstein's equation when eqs~(\ref{Einstein:tt})-(\ref{eq:Poisson}) 
hold, provided, of course, that conditions (\ref{condi:Newtonian}) 
are satisfied. 

We now assert that the metric, eq.~(\ref{metric:Newton-FLRW}), appears
to very accurately describe our universe on {\em all scales}, except
in the immediate vicinity of black holes and neutron stars.  The basis
for this assertion is simply that the FLRW metric appears to provide a
very accurate description of all phenomena observed on large scales,
whereas Newtonian gravity appears to provide an accurate description
of all phenomena observed on small scales.  The metric
(\ref{metric:Newton-FLRW}) together with eqs~(\ref{condi:Newtonian}), 
predicts that large 
scale phenomena will be accurately described by a FLRW model, whereas
the metric (\ref{metric:Newton-FLRW}) together with
eq.~(\ref{eq:Poisson}) predicts that Newtonian gravity will hold 
in regions small compared with the Hubble radius 
in which the stress-energy (\ref{Tm}) dominates over (\ref{Ts})
(see \cite{HolzWald97}).

Note that the validity of eq.~(\ref{metric:Newton-FLRW}) for accurately
describing phenomena on {\it small} scales holds despite the fact that
the density contrast of matter is commonly quite 
large~\cite{Barrow88,Futamase89} 
\ben
 \frac{\delta \rho}{\rho} \gg 1 \,.
\een 
Indeed, for the solar system, galaxies, clusters of galaxies, 
we can estimate respectively, 
$\delta \rho / \rho \approx 10^{30},\, \approx 10^{5},\, \approx 10^{2} 
\gg 1$. Nevertheless, in all of these cases, we have
$\Psi \approx 10^{-6} \sim 10^{-5} \ll 1$, and the other conditions 
appearing in eq.~(\ref{condi:Newtonian}) also hold.
Even for neutron stars, $\Psi \approx 10^{-1}$, 
so the metric eq.~(\ref{metric:Newton-FLRW}) is probably not too bad an 
approximation even in the vicinity of neutron stars.

The key point of this section is that 
if our assertion is correct that a metric of the form of 
eq.~(\ref{metric:Newton-FLRW}) accurately describes our universe,
and if it also is true that 
conditions (\ref{condi:Newtonian}) hold, then the nonlinear correction 
terms occurring in eqs.~(\ref{Einstein:tt}) and (\ref{Einstein:ij}) 
are negligibly small. It therefore is manifest that nonlinear 
corrections\footnote{ 
We take this opportunity to comment upon one misconception 
related to the validity of the Newtonianly perturbed FLRW metric. 
It is commonly stated that when $\delta \rho/{\rho} \gg 1$, one enters 
a ``nonlinear regime.'' This might suggest that the validity of 
eqs.~(\ref{metric:Newton-FLRW}) and/or (\ref{eq:Poisson}) 
would be questionable whenever $\delta \rho/{\rho} \gg 1$. 
However, this is not the case; the proper criteria for the validity 
of the metric eq.~(\ref{metric:Newton-FLRW}) are conditions 
(\ref{condi:Newtonian}), not $\delta \rho/{\rho} \ll 1$. 
It is true that nonlinear effects become important for the motion of 
matter when $\delta \rho/{\rho} \gg 1$. 
This follows simply from the fact that self-gravitation is a nonlinear effect, 
and self-gravitational effects on the motion of matter cannot be ignored 
when $\delta \rho/{\rho} > 1$. 
But this does not mean that one must include nonlinear corrections 
to the metric form, eq.~(\ref{metric:Newton-FLRW}), or to 
eq.~(\ref{eq:Poisson}) in order to get a good approximation to the
spacetime metric. Indeed, if one is trying to describe the solar
system in the context of the ordinary Newtonian limit of general
relativity, one must include ``nonlinear effects'' to obtain the
correct motion of the planets; they would move on geodesics of the
flat metric rather than the Newtonianly perturbed metric if not for
these ``nonlinear effects.'' However, the corrections to the spacetime 
metric of the solar system arising from nonlinear terms in Einstein's
equation are entirely negligible.
} 
to the dynamics of the universe will be negligible, 
i.e., there will be no important ``back-reaction'' effects of 
the inhomogeneities on the observed expansion of the universe on large scales. 
In particular, accelerated expansion cannot occur if the smoothly distributed 
matter satisfies the strong energy condition. 
However, our assertion that the metric, eq.~(\ref{metric:Newton-FLRW}), 
very accurately describes our universe is merely an assertion, 
and we cannot preclude the possibility that other models 
(e.g., with large amplitude, long-wavelength gravitational waves 
or with matter density inhomogeneities of a different type) might also 
fit observations. Our main point of this paper, however, is that 
if one wishes to propose an alternative model, then it is necessary 
to show that all of the predictions of this model are compatible 
with observations such as the observed redshift-luminosity relation for 
type Ia supernovae and the various observed properties of the cosmological 
microwave background (CMB) radiation. 
As we shall illustrate in the next two sections, 
it does not suffice to show merely that the spatially 
averaged scale factor behaves in a desired way or that an effective 
stress-energy tensor is of a desired form.

\section{Cosmic Acceleration via Averaging} 

The type of spatial averaging in the context of Newtonianly perturbed
FLRW models that was done to derive eqs.~(\ref{Einstein:tt}) and 
(\ref{Einstein:ij}) above is not problematical. 
The metric very nearly has FLRW symmetry, so there 
is a natural choice of spatial slices on which one can take spatial
averages. Since $\Psi \ll 1$, it makes negligible difference if one
uses the spacetime metric (\ref{metric:Newton-FLRW}) or the
corresponding ``background'' FLRW metric (i.e.,
eq.~(\ref{metric:Newton-FLRW}) with $\Psi$ set equal to zero) to
define the averaging.

If one has a metric that does not nearly have FLRW symmetry, one can,
of course, still define spatial averaging procedures. However, these
will, in general, be highly dependent on the choice of spatial slicing, 
and the results obtained from spatial averaging need not be interpretable 
in a straightforward manner. We now illustrate these comments with concrete 
examples.

For simplicity and definiteness, we consider an inhomogeneous universe
with irrotational dust.  In the comoving synchronous gauge,\footnote{ 
If the universe is filled with irrotational dust, then the comoving 
synchronous gauge defines a natural choice of slicing, namely the slices 
orthogonal to the world lines of the dust. However, for an inhomogeneous 
universe, this gauge choice typically will break down on timescales much 
shorter than cosmological timescales, due to formation of caustics. 
For example, synchronous coordinates defined in a neighborhood of the Earth 
would typically break down on a timescale of order the free fall time to the 
center of the Earth, i.e., $\sim 1$ hour. 
 } 
the metric takes the form 
\ben 
ds^2 = -d t^2 + q_{ij}(t,x^m) dx^idx^j \,.  
\een
Let $\Sigma$ denote a hypersurface of constant $t$, 
let $\D$ denote a compact region of $\Sigma$ 
and let $\phi$ be a scalar field on
$\Sigma$.  The average, $\langle \phi \rangle_\D$, of $\phi$ over the
domain $\D \subset \Sigma$ may be defined by
\ben 
\langle \phi \rangle_\D \equiv \frac{1}{V_\D} \int_\D \phi d\Sigma \,,
\label{def:averaging}
\een 
where $V_\D$ denotes the volume of $\D$ 
and $d\Sigma$ is the proper volume element of $\Sigma$. 
\footnote{ 
For a different type of averaging procedure than that given by 
eq.~(\ref{def:averaging}) and its application to cosmology, 
see e.g.,~\cite{Zalaletdinov97,CPZ05}. } 
Define the averaged scale factor, $a_\D$, by 
\ben
a_\D \equiv (V_\D)^{1/3} \,. 
\een
We ``time evolve'' $\D$ by making it be comoving with the dust, 
i.e., the (comoving) coordinates of the boundary of $\D$ remain constant 
with time. Following Buchert~\cite{Buchert00}, one then obtains 
from Einstein's equation the following equations of motion for $a_\D$,  
\bena
  3\frac{\ddot a_\D}{a_\D} 
  &=& -\frac{\kappa^2}{2} \la \rho \ra_\D + Q_\D \,, 
\label{eq:evolve}
\\ 
  3 \left( \frac{\dot a_\D}{a_\D} \right)^2 
  &=& \kappa^2\la \rho \ra_\D  -\half \la {\cal R} \ra_\D - \half Q_\D \,,
\label{eq:friedman}
\eena
together with
\ben
 \Tdot{\left( a_\D^6 Q_\D \right)}  
 + a_\D^4 
 \Tdot{\left(a_\D^2 \la {\cal R} \ra_\D \right)} =0 \,. 
\label{condi:integrability} 
\een
Here ${\cal R}$ is the scalar curvature of $\Sigma$ and 
\bena
  Q_\D &\equiv& \frac{2}{3} \left( \langle \theta^2 \rangle_\D 
               - \langle \theta \rangle_\D^2 \right)
               - \langle \sigma_{ij}\sigma^{ij} \rangle_\D \,,  
\eena  
In deriving these equations, it is important to 
bear in mind that the averaging of the time derivative
$\langle {\dot \phi} \rangle_\D \equiv 
\langle { \partial \phi}/\partial t \rangle_\D$ 
of a locally defined quantity $\phi$ differs in general from 
the time derivative of the averaged quantity 
$\Tdot{\langle \phi \rangle}_\D \equiv 
\partial {\langle \phi \rangle}_\D/\partial t$, 
since the volume element and $V_\D$ may depend on the time $t$.  
Indeed, we have
\ben
\Tdot{\langle \phi \rangle}_\D 
= \langle {\dot \phi} \rangle_\D + \langle \theta \phi \rangle_\D 
- \langle \theta \rangle_\D \langle \phi \rangle_\D \,, 
\een
where $\theta$ denotes the expansion of the world lines of the dust.

It is immediately seen from (\ref{eq:evolve}) that ``averaged acceleration'' 
$\ddot a_\D>0$ is achieved if 
\ben
 Q_\D  > \frac{\kappa^2 }{2} \la \rho \ra_\D \,.  
\label{condi:acceleration}
\een 
Buchert~\cite{Buchert05} has discussed 
cosmological implications of the condition (\ref{condi:acceleration}).

A number of authors (see, e.g., Refs.~\cite{KMR05, NT05, Moffat05}) 
have sought to account for the observed
acceleration of our universe by means of inhomogeneous models that
satisfy eq.~(\ref{condi:acceleration}). However, our main point of this 
section is that even if our universe (or a suitable spatial domain of our 
universe) satisfies eq.~(\ref{condi:acceleration}) and thus has $\ddot
a_\D>0$, this does {\it not} imply that the model will possess any
physically observable attributes of an accelerating FLRW model. Indeed
Nambu and Tanimoto~\cite{NT05} have shown that in a cosmological model in 
which $\D$ can be written as a union of regions each of which is locally 
homogeneous and isotropic, we have
\ben
 a_\D^2\ddot{a_\D} = a_1^2{\ddot a}_1  + a_2^2{\ddot a}_2 + \cdots   
                   + \frac{2}{a_\D^3 } 
                            \sum_{i\neq j} a_i^3a_j^3 
                            \left(
                                  \frac{\dot a_i}{a_i}- \frac{\dot a_j}{a_j} 
                            \right)^2 \,,  
\label{nt}
\een
where $a_i$ denotes the locally defined scale factor in the $i$-th
patch, and, in this case, $a_\D \equiv (a_1^3 + a_2^3 + 
\cdots)^{1/3}$.  Consider, now, a model where at time $t$ the universe
consists of two disconnected(!) dust filled FLRW models, one of which is 
expanding and the other of which is contracting. Both components of the 
universe are, of course, decelerating, i.e., ${\ddot a}_1 < 0$, 
${\ddot a}_2 < 0$. Nevertheless, it is not difficult to see from 
eq.~(\ref{nt}) that ${\ddot a}_\D > 0$ can easily be satisfied. 
For example, if we take $a \equiv a_1 = a_2 $ and 
$\dot{a}_1 = - \dot{a}_2$ we obtain 
\ben 
 a_\D^2 {\ddot a}_\D  
  = 
 2a^3 \left\{\frac{\ddot a}{a} + 4\left(\frac{\dot a}{a}\right)^2 \right\} 
  = \frac{7}{3} \kappa^2 a^3 \rho > 0 \,.
\een
We thereby obtain a very simple model where the universe is 
accelerating according to the definition eq.~(\ref{condi:acceleration}), 
but all observers see only deceleration. This graphically illustrates that 
satisfaction of eq.~(\ref{condi:acceleration}) in a model is far from 
sufficient to account for the physically observed effects of 
acceleration in our universe. 
We see no reason to believe that the spatially averaged acceleration 
found, e.g., in the models of \cite{NT05,Moffat05} directly corresponds 
to any physical effects of acceleration such as would be observed in 
type Ia supernovae data. The only way to tell if a model displays physically 
observable effects of acceleration is to calculate these effects. 

As already mentioned above, the averaging procedure defined by 
eq.~(\ref{def:averaging}) also has ambiguities both with regard to 
the choice of time slicing and the choice of domain $\D$. 
One may also artificially produce an
averaged cosmic acceleration as a result of a suitably chosen time-slicing. 
To show this explicitly, we
give an example of accelerated expansion in Minkowski spacetime.
We note first that for a general inhomogeneous universe 
(i.e., with no assumption concerning the form of the stress-energy), 
the equation of motion for $a_\D$ can be expressed as 
\bena  
 3\frac{\ddot a_\D}{a_\D} 
 = - \langle {\cal R} \rangle_\D 
 - 6\left(\frac{\dot a_\D}{a_\D} \right)^2 
 + \langle (G_{ab} + \frac{1}{2}g_{ab}G^c{}_c)t^at^b \rangle_\D \,,  
\label{average:acceleration}   
\eena 
where $G_{ab}$ denotes the Einstein tensor.   
Therefore, for any vacuum spacetime, if there is a domain $\D$ such that 
$- \langle {\cal R} \rangle_\D > 6\left({\dot a_\D}/{a_\D} \right)^2 $, 
then $\D$ describes an accelerated expansion insofar as $a_\D$ is concerned.

\begin{figure}[h] 
 \centerline{\epsfxsize = 7cm \epsfbox{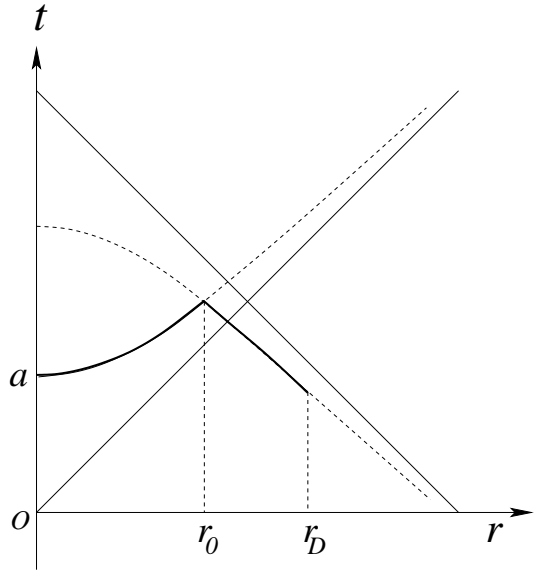}}  
\vspace{3mm}
\begin{center}
\begin{minipage}{12cm}
     \caption{\small 
             An accelerating domain ${\cal D}$ (thick line) in Minkowski 
             spacetime $(t,r)$ with the angular coordinates suppressed. 
             The domain $\D$ is constructed by cutting portions out of 
             the two hyperboloids (thin dashed lines), 
             $t = \sqrt{a^2 + r^2}$ and 
             $t = 2\sqrt{a^2 + r_0^2} - \sqrt{a^2 + r^2}$,  
             and joining them at $r=r_0$. 
             The matching can be done in a smooth manner, 
             as explained in Appendix. 
             The boundary radius $r=r_\D$ is chosen so that $\dot a_\D =0$. 
             } 
        \protect \label{fig:Accele-slicing} 
\end{minipage}
\end{center}
\end{figure}

To construct an accelerating $\D$ in Minkowski spacetime, we start with 
two hyperboloidal slices, one of which corresponds to an expanding time-slice 
in the Milne chart covering the future of the origin 
\ben
 ds^2 = -da^2 + a^2 (d\xi^2 + \sinh^2 \xi d\Omega^2) \,, 
\een
and the other of which is a similar hyperboloidal slice that is contracting.
The idea of the construction is to join these two hyperboloids at some
radius, smooth out the join region, and choose $\D$ so that ${\dot a}_\D = 0$
(see Figure \ref{fig:Accele-slicing}). 
Since the hyperboloids have negative scalar curvature, we thereby have 
${\cal R} < 0$ except near the radius where the hyperboloids are joined.
However, we can show that this construction can be done so that
the contribution from the join region can be made arbitrarily small. 
Consequently, we obtain $3{\ddot a_\D}/{a_\D} 
= - \langle {\cal R} \rangle_\D > 0$. Details of this 
construction are given in the appendix. Since Minkowski spacetime 
does not display
any physical effects associated with accelerated expansion, 
this example shows quite 
graphically that ``acceleration'' as defined by the 
above averaging procedure can 
easily arise as a gauge artifact produced by a suitable choice of time slicing.

\section{Cosmological Back-reaction in the Long-wavelength Limit} 

A number of authors have considered the back-reaction effects of
perturbations of FLRW models, particularly with regard to modes whose
wavelength is comparable to or larger than the Hubble 
radius~\cite{MAB1997Lett,ABM1997PRD,Nambu02,BL2004}. 
The basic strategy has been to compute the second order 
terms in Einstein's equation arising from these perturbations, thereby
obtaining an ``effective stress-energy tensor'' for the
perturbations. If the form of this effective stress-energy tensor
corresponds to that of a positive cosmological constant of the correct
magnitude, then one might hope to have provided a mechanism for obtaining
the observed acceleration of our universe as a back-reaction effect of
long-wavelength perturbations, without the need to introduce a
cosmological constant or dark energy.\footnote{ 
The motivation in~\cite{MAB1997Lett,ABM1997PRD} was actually 
to use back-reaction effects to attempt to cancel the presence of a 
large cosmological constant rather than to use back-reaction to 
directly produce acceleration.  
} 

We comment, first, that, even without extensive analysis, there is an
intrinsic implausibility to this type of explanation.  If
one considers perturbations of wavelength less than the Hubble radius,
it is hard to imagine that the perturbations could be so small that we
do not notice any significant deviations from homogeneity and
isotropy, yet so large that their {\it second order} effects produce
very significant changes to the dynamics of our universe. On the other
hand, if one goes to the long-wavelength limit, then the
perturbation should correspond closely to a perturbation to a spatially
homogeneous cosmological model. But, given the severe constraints on
anisotropy arising from CMB observations, the perturbation should, in fact,
correspond to a perturbation towards another FLRW model. Thus, one
should not be able to obtain any new phenomena (such as acceleration
without a cosmological constant or dark energy)
that are not already present in FLRW models.

At least part of the confusion with regard to the calculation of the
back-reaction effects of cosmological perturbations appears to stem
from the fact that the notion of an ``effective stress-energy tensor''
for perturbations arises in two quite different contexts, namely (i)
ordinary perturbation theory and (ii) the ``shortwave approximation.''
We now explain this distinction. For simplicity, we restrict
consideration in the following discussion to the vacuum case;
cosmological perturbations for the Einstein-scalar-field system will
be considered later in this section.

Ordinary perturbation theory (see, e.g., section 7.5 of \cite{w84}) arises by
considering a one-parameter family of metrics $g_{ab}(\alpha)$ that is
jointly analytic in its dependence on $\alpha$ and the spacetime point.
We refer to $g^{(0)}_{ab} \equiv g_{ab}(0)$ as the
``background metric.'' Roughly speaking, as $\alpha \rightarrow 0$,
$g_{ab}(\alpha)$ differs from $g^{(0)}_{ab}$ by a perturbation that becomes
of arbitrarily small amplitude but maintains a fixed profile. One expands
$g_{ab}(\alpha)$ in a power series in $\alpha$ about $\alpha = 0$
\ben
g_{ab}(\alpha) = \sum_n \frac{1}{n!} \alpha^n g^{(n)}_{ab} \,. 
\een
The perturbation equations for $g^{(n)}_{ab}$ for the vacuum 
Einstein equation, 
\ben
G_{ab} = 0 \,, 
\label{ein}
\een
are then obtained by differentiating the Einstein tensor, 
$G_{ab} (\alpha)$, of $g_{ab}(\alpha)$ $n$ times with respect to 
$\alpha$ at $\alpha = 0$. 
The zeroth order equation is just Einstein's equation for $g^{(0)}_{ab}$ 
\ben
G_{ab} [g^{(0)}] = 0 \,. 
\label{0th}
\een
The first order equation is
\ben
G^{(1)}_{ab} [g^{(1)}] = 0 \,, 
\label{1st}
\een
where $G^{(1)}_{ab}$ denotes the linearized Einstein tensor off 
of the background metric $g^{(0)}_{ab}$. The second order equation is 
\ben
G^{(1)}_{ab} [g^{(2)}] = - G^{(2)}_{ab} [g^{(1)}] \,, 
\label{2nd}
\een
where $G^{(2)}_{ab}[g^{(1)}]$ denotes the second-order Einstein tensor 
constructed from $g^{(1)}_{ab}$. 

As can be seen from eq.~(\ref{2nd}), minus the second-order Einstein
tensor (divided with the gravitational constant), 
$- \kappa^{-2}G^{(2)}_{ab} [g^{(1)}]$, plays the role of an
``effective stress-energy tensor'' associated with the perturbation 
$g^{(1)}_{ab}$ in the sense that it acts as a source term for 
the second-order metric perturbation $g^{(2)}_{ab}$. 
However, this does {\it not} mean that one can treat 
$- \kappa^{-2}G^{(2)}_{ab} [g^{(1)}]$ as though it were a new form of matter 
stress-energy that can be inserted into the right side of the exact 
Einstein equation (\ref{ein}) as opposed to the right side of 
eq.~(\ref{2nd}). For one thing, the second-order Einstein tensor is 
highly gauge dependent (as we shall illustrate explicitly below), so it is 
not straightforward to interpret its meaning. 
The key point, however, is that eq.~(\ref{2nd}) arises only in the context of 
perturbation theory. If $G^{(2)}_{ab} [g^{(1)}]$ is very small, then 
its effects on the spacetime metric can be reliably calculated from 
eq.~(\ref{2nd}). But if $G^{(2)}_{ab} [g^{(1)}]$ is large enough to produce 
cosmologically interesting effects (such as acceleration), then the third 
and higher order contributions to $g_{ab}(\alpha)$ will also be large, 
and one cannot reliably compute back-reaction effects from second-order 
perturbation theory. 

This situation occurring in perturbation theory contrasts sharply with
the situation that arises when one uses the ``shortwave
approximation'' \cite{Isaacson1968, MTW, Burnett1989}. 
Here, one wishes to develop a formalism in which the self-gravitating 
effects of gravitational radiation---and the consequent effects on 
the spacetime metric on scales much larger than the wavelength of the
radiation---can be reliably obtained, even when these effects are
``large.'' Again, one considers a one-parameter family of
metrics $g_{ab}(\beta)$ that has a continuous limit to the metric
$g^{(0)}_{ab} \equiv g_{ab}(0)$. Thus, as in ordinary perturbation
theory, as $\beta \rightarrow 0$, $g_{ab}(\beta)$ differs from
$g^{(0)}_{ab}$ by a perturbation of arbitrarily small amplitude.
However, one now requires $g_{ab}(\beta)$ to be such that, roughly
speaking, as $\beta \rightarrow 0$, the ratio of the
amplitude to the wavelength of the perturbation goes to a finite, non-zero 
limit; see \cite{Burnett1989} for a precise statement of what is 
required in this limit. Thus, in this scheme,
the dominant terms in Einstein's equation as $\beta \rightarrow 0$ are 
actually the linear terms in the second derivatives of
the first order perturbation, which diverge as $1/\beta$. One thereby obtains
\ben
G^{(1)}_{ab} [g^{(1)}_{ab}] = 0 \,. 
\een
The quadratic terms in the first order perturbation are of zeroth order in 
$\beta$, so they make a contribution to the Einstein tensor that is 
comparable to that of $g^{(0)}_{ab}$. The linear terms in the second-order 
perturbation also contribute to this order, but these contributions can be 
eliminated by averaging over a spacetime region that is large compared with 
the wavelength of the perturbation. One thereby obtains 
\ben
G_{ab} [g^{(0)}] = \langle - G^{(2)}_{ab} [g^{(1)}] \rangle \,, 
\label{2ndsw}
\een
where the brackets on the right side of eq.~(\ref{2ndsw}) denote a 
suitably defined spacetime average. It can be shown that 
$\langle - G^{(2)}_{ab} [g^{(1)}] \rangle$ is gauge invariant in a suitably 
defined sense. We refer to \cite{Burnett1989} for further details of 
the derivation and meaning of these equations.

Although eq.~(\ref{2ndsw}) is quite similar in form to eq.~(\ref{2nd}), 
the meaning and range of validity of these equations are quite different. 
In contrast to eq.~(\ref{2nd}), it should be possible to 
use eq.~(\ref{2ndsw}) to calculate the back-reaction effects 
of gravitational radiation even when these effects are large. 
The catch, however, is that eq.~(\ref{2ndsw}) can be used only when 
the wavelength of the perturbation is much smaller than the curvature 
lengthscale of the background spacetime. Thus, if the universe were filled 
with gravitational radiation of wavelength much smaller than the Hubble 
radius, then it should be possible to use eq.~(\ref{2ndsw}) to 
reliably calculate the back-reaction effects of this radiation, even if
this radiation is the dominant form of ``matter'' in the universe. 
However, eq.~(\ref{2ndsw}) manifestly {\it cannot} be used to calculate 
the back-reaction effects of long-wavelength perturbations. 

We conclude this section by deriving an explicit formula for the gauge
dependence of the second-order ``effective stress-energy tensor'' 
arising in ordinary perturbation theory
for long-wavelength scalar-type perturbations of a FLRW universe containing 
a scalar field. 
By doing so, we will see that one can get essentially any answer 
one wishes for this effective stress-energy tensor by making appropriate 
gauge transformations. This graphically shows that one cannot draw 
any physical conclusions merely by examining the form of the effective 
stress-energy tensor arising in second-order perturbation theory.
\footnote{ 
Discussion of the characterization of the back-reaction effects 
of perturbations in terms of physical variables can be found 
in \cite{AW01,GB05}. } 

Consider a background flat FLRW universe 
\ben
  ds^2 = - dt^2 + a^2(t) \gamma_{ij}dx^idx^j \,, 
\een
filled with a scalar field whose energy momentum tensor is given by 
\ben
 T_{ab } = \nabla_a \phi \nabla_b \phi 
 - \frac{1}{2} g_{ab } \{\nabla^c\phi \nabla_c \phi + 2U(\phi) \} \,.   
\label{stress:scalar}
\een
The unperturbed background equations of motion for $a$ and $\phi$, 
which are functions of only $t$, are given by 
\bena
&& H^2 \equiv \left(\frac{\dot a}{a}\right)^2 
   = \frac{\kappa^2}{3}\left(\frac{1}{2}\dot{\phi}^2+ U \right) \,, 
\\
&& 
  \ddot \phi + 3H \dot \phi + \frac{\partial U}{\partial \phi} = 0 \,,  
\eena
where in these equations and hereafter the {\it dot} denotes 
the derivative with respect to $t$.


We focus on the scalar-type perturbations. The general form of a scalar-type
metric perturbation is 
\ben
d \tilde{s}^2 = - (1 + 2A\scalar )dt^2 - 2 a B\scalar_i dt dx^i 
          + a^2 \{ (1 + 2H_L\scalar )\gamma_{ij} 
          + 2H_T\scalar_{ij} \} dx^idx^j \,,  
\label{metric:pert}
\een
and the scalar field perturbation is given by 
\ben
  \tilde \phi = \phi + \delta \phi \scalar \,. 
\een
Here $\scalar$ denotes a plane wave  
on flat $3$-space with wavevector ${\bf k}$, and 
$\scalar_{i}$ and $\scalar_{ij}$ are the divergence-free vector 
and transverse-traceless tensor defined by 
\ben
  \scalar_{i} = - \frac{1}{k}D_i\scalar \,,\quad 
  \scalar_{ij} = \frac{1}{k^2}
                 \left( 
                       D_iD_j - \frac{1}{3}\gamma_{ij}\Delta_{(3)} 
                 \right)\scalar \,,  
\een
with $D_i$ being the derivative operator associated with the $3$-space 
metric $\gamma_{ij}$ as in (\ref{condi:Newtonian}), 
and $k^2={\bf k}\cdot {\bf k}$. 
Here and in the following, perturbation variables are understood as 
corresponding Fourier expansion coefficients---hence functions 
merely of $t$---and we omit the index ${\bf k}$ unless otherwise stated. 

Under infinitesimal gauge transformations of the scalar-type;   
\ben
 t \rightarrow t +   T\scalar \,, \quad 
 x^i \rightarrow x^i + L \scalar^i \,,    
\label{coord}
\een 
the perturbation variables, $A,\,B,\,H_L,\,H_T,\, \delta \phi$, change as 
\bena 
 A  &\rightarrow&  A - \dot T \,,         \label{gt:A}\\
 B  &\rightarrow&  B + a\dot L
                     + \frac{k}{a}{T} \,, \label{gt:B}  \\ 
 H_L &\rightarrow& H_L - \frac{k}{3} {L} 
                       - H {T} \,,        \label{gt:HL} \\
 H_T &\rightarrow& H_T + k{L} \,,         \label{gt:HT} \\ 
 \delta \phi &\rightarrow& \delta \phi - \dot \phi {T} \,. \label{gt:phi}
\eena
In particular, it follows from eqs.~(\ref{gt:B}) and (\ref{gt:HT}) that 
the following combinations 
\ben
  X_T \equiv \frac{a}{k} \left( \frac{a}{k}\dot{H_T} - B \right) \,,   
 \quad 
  X_L \equiv - \frac{1}{k}H_T \,,     
\een 
change as 
\ben
  X_T \rightarrow X_T -  T \,, \quad 
  X_L \rightarrow X_L -  L \,.     
\label{gaugetransf:XTXL}
\een
Hence, by inspection of eqs.~(\ref{gt:A}), (\ref{gt:HL}) and (\ref{gt:phi}), 
one can immediately obtain gauge-invariant perturbation 
variables \cite{Bardeen1980,KS1984};  
\ben
 \Psi \equiv A - \dot{X_T} \,, \quad 
 \Phi \equiv H_L - \frac{k}{3}X_L - HX_T \,, \quad 
 \Dphi \equiv \delta \phi - \dot \phi X_T \,.    
\een
Any scalar-type gauge-invariant perturbation quantity can be expressed as 
a linear combinations of the gauge-invariant variables 
$\Psi$, $\Phi$, and $\Dphi$, and their time derivatives.

It follows from the linearized Einstein equations that 
the gauge-invariant variables defined above satisfy the following 
equations~\cite{KS1984} 
\bena 
  \frac{k^2}{a^2}\left( 
                       \Psi + \Phi 
                 \right) = 0 \,,  
\label{Ein:ij:tracefree} 
\qquad 
   k \left[ 
           2(\dot \Phi - H \Psi) + \kappa^2 \dot \phi \Dphi 
     \right] = 0 \,.       
\label{Ein:0j}
\eena 
which correspond, respectively, to the trace-free part of 
the space-space component and the time-space component of 
the linearized Einstein equations. 
For $k^2 \neq 0$, we obtain from eq.~(\ref{Ein:0j}) 
the following relations between $\Phi$, $\Psi$, and $\Dphi$;  
\ben
 \Phi= - {\Psi}   \,, 
    \quad 
 \frac{\Dphi}{\dot \phi}
  = - \frac{1}{\dot H} \left( \dot \Psi + H \Psi \right) \,.   
\label{rel:Psi:Phi:X}
\een 
It then also follows from Einstein equations that $\Psi$ is governed by 
\bena
&& \ddot \Psi 
  + \left(
          H - \frac{{\ddot H}}{{\dot H}} 
    \right)\dot \Psi 
  + \left( 
            2\frac{\dot H}{H} - \frac{{\ddot H}}{{\dot H}} 
    \right) H \Psi 
  + \frac{k^2}{a^2} \Psi = 0 \,.  
\label{eom:Psi} 
\eena

We therefore have found that all of the scalar-type 
perturbations variables are 
given in terms of the variables $\Psi$, $X_T$, and $X_L$ by 
\bena
 A &=& \Psi + \dot{X_T} \,, 
\label{expr:A}
\\
 B &=& - a \dot{X_L} - \frac{k}{a}X_T \,, 
\label{expr:B}
\\ 
 H_L &=& 
        - {\Psi} + H X_T + \frac{k}{3}X_L \,, 
\label{expr:HL}
\\  
 H_T &=& - k X_L \,, 
\label{expr:HT}
\\
 \frac{\delta \phi}{\dot \phi} 
   &=&     - \frac{1}{{\dot H}} 
             \left( {\dot \Psi} + H \Psi\right) 
           + X_T \,. 
\label{expr:Delta}
\eena
The variable $\Psi$ is gauge invariant and satisfies eq.~(\ref{eom:Psi}).
On the other hand, the gauge transformation law (\ref{gaugetransf:XTXL}) 
implies 
that the functions $X_T$ and $X_L$ are completely arbitrary, 
i.e., they may be chosen to take any values that one wishes.
Thus, the specification of $X_T$ and $X_L$ in terms of $\Psi$ corresponds 
to fixing the gauge freedom.  
For example, the choice  
\ben
 X_T=X_L=0 
\een 
corresponds to the Poisson gauge (or the longitudinal gauge), 
in which the metric, eq.~(\ref{metric:pert}), takes precisely 
the form of eq.~(\ref{metric:Newton-FLRW}). 
Another example is the choice  
\bena
 X_T &=& - \int^t_{t_*} \Psi(t') dt'+ C_1(k) \,, 
\label{X:gauge:synchronous:1}
\\
 X_L &=&  k\int^t_{t_*} 
           \left\{\int^{t'}_{t'_*} \Psi(t'') dt'' \right\}\frac{dt'}{a^2(t')} 
            - k C_1(k) \int^t_{t_*} \frac{dt'}{a^2(t')} 
            + C_2(k) \,,  
\label{X:gauge:synchronous:2}
\eena
where $t_*$ denotes some reference time and $C_1$ and $C_2$ are 
arbitrary constants. This choice corresponds to the synchronous 
gauge, $A=B=0$, in which $C_1$ and $C_2$ parameterize the residual 
gauge freedom in this gauge. 

The second-order effective stress-energy tensor for the Einstein-scalar-field
system is defined by
\ben
  {}^{({\rm eff})}\!T_{ab} \equiv
  - \frac{1}{\kappa^2} G^{(2)}_{ab}[g^{(1)}] 
  + T^{(2)}_{ab}[\delta \phi,g^{(1)}] \,, 
\een
where $G^{(2)}_{ab}$ denotes the second order Einstein tensor and
$T^{(2)}_{ab}$ is the similarly defined second order contribution to $T_{ab}$,
eq.~(\ref{stress:scalar}), arising from the first order perturbation
$(\delta \phi,g^{(1)})$.
We now calculate the second-order effective stress-energy tensor
in order to explicitly
demonstrate its gauge dependence. It is very convenient to 
express ${}^{({\rm eff})}\!T_{ab}$ in terms
of the variables $\Psi$, $X_T$, and $X_L$, since any 
dependence of ${}^{({\rm eff})}\!T_{ab}$ on $X_T$, and $X_L$ will explicitly
show its gauge dependence. Clearly, since ${}^{({\rm eff})}\!T_{ab}$ 
is quadratic in the first order perturbation, it must consist
of a part, ${}^{({\rm eff:}\Psi)}\!T_{ab}$, that is quadratic in $\Psi$, 
a part, ${}^{({\rm eff:}X)}\!T_{ab}$, that is quadratic in $(X_T, X_L)$,
and a part, ${}^{({\rm eff:}\Psi,X)}\!T_{ab}$, containing the ``cross-terms''
between $\Psi$ and $X_T, X_L$. The quantity 
${}^{({\rm eff:}\Psi)}\!T_{ab}$ is gauge invariant, 
\footnote{ 
In fact, ${}^{({\rm eff:}\Psi)}\!T_{ab}$ is precisely the 
``effective energy-momentum tensor for cosmological perturbations'' 
of \cite{MAB1997Lett,ABM1997PRD}. 
However, contrary to the claims of \cite{MAB1997Lett,ABM1997PRD}, 
this effective energy-momentum tensor is 
gauge-invariant only in the trivial sense that any gauge-dependent 
quantity can be viewed as gauge invariant once a gauge has been completely 
fixed. In the variations taken in \cite{MAB1997Lett,ABM1997PRD} to obtain 
their effective energy-momentum tensor, $X_T$ and $X_L$ were implicitly 
assumed to be independent of $\Psi$, corresponding to the choice of 
the Poisson (longitudinal) gauge $X_T=X_L=0$. However, 
different specifications of $X_T$ and $X_L$ in terms of $\Psi$---i.e., 
different choices of gauge---would lead to {\em different} expressions  
for the effective energy-momentum tensor in terms of $\Psi$. 
 } 
but both ${}^{({\rm eff:}X)}\!T_{ab}$ and ${}^{({\rm eff:}\Psi,X)}\!T_{ab}$
are gauge dependent. Thus, ${}^{({\rm eff})}\!T_{ab}$ will be gauge invariant
if and only if these latter pieces vanish.

It is easy to verify that both ${}^{({\rm eff:}X)}\!T_{ab}$ 
and ${}^{({\rm eff:}\Psi,X)}\!T_{ab}$ are nonvanishing. 
To see explicitly that ${}^{({\rm eff:}X)}\!T_{ab}$ is nonvanishing, 
it suffices to consider the case where we impose the additional restriction
\ben
 X_L = - k\int^t_{t_*} \frac{X_T(t') }{a^2(t')} dt' \,, 
\een  
which ensures that $B=0$, thereby considerably simplifying the calculation.
We also focus attention on the long-wavelength limit.
A brute force calculation then yields
\bena
 \kappa^2 \, {}^{({\rm eff:}X)}T_{00} 
 &=& 
 \Bigg[
       - ({\dot H} + 3H^2){\dot X_T}^2 
       + ({\ddot H}+ 6H{\dot H} + 12H^3)X_T{\dot X_T}  
\non \\  
 && \quad 
     + \left( 3{\dot H}^2 + 12{\dot H}H^2+ 3H{\ddot H} + {\dddot H} 
             -\frac{3}{4} \frac{{\ddot H}^2}{\dot H} 
       \right) X_T^2 
 \Bigg] + O(k^2)\,,
\label{T00} 
\\
 \kappa^2 \, {}^{({\rm eff:}X)}T_{ij} 
 &=& g_{ij}
 \Bigg[
       2H{\dot X_T}{\ddot X_T} -2({\dot H} + 2H^2)X{\ddot X} 
       - ( {\dot H} + H^2 ) {\dot X_T}^2 
\non \\
 && \qquad 
       - ( 3{\ddot H} + 22 H{\dot H} + 12H^3 )X_T {\dot X_T} 
\non \\
 && \qquad 
       -\left(
               \frac{11}{2}H{\ddot H} + 4{\dot H}^2 
              + 12{\dot H}H^2 + \half {\dddot H} 
        \right) X_T^2 
 \Bigg] + O(k^2)\,. 
\label{Tij}
\eena 
Thus, even in the case of a pure gauge perturbation, $\Psi = 0$, we can 
obtain a non-vanishing effective stress-energy tensor for 
long-wavelength perturbations. Indeed, since $X_T$ is 
entirely arbitrary, we see that we can get essentially any answer one wishes
for ${}^{({\rm eff})}\!T_{ab}$. For example, if one wishes to have a pure
gauge perturbation in which ${}^{({\rm eff})}\!T_{ab}$ takes the form of a
cosmological constant, one would merely have to solve the second-order 
ordinary differential equation for $X_T$ that results when one equates the
right side of eq.~(\ref{T00}) to minus the right side of eq.~(\ref{Tij}).
This manifestly demonstrates that one cannot derive any physical consequences
by merely examining the form of the second-order effective stress-energy 
tensor.

Finally, we comment that we derived the above ``long-wavelength limit'' 
form of ${}^{({\rm eff})}\!T_{ab}$ for scalar-type perturbations by 
considering perturbations with $k^2 \neq 0$ and then taking the limit 
as $k^2 \rightarrow 0$. Alternatively, we could have directly 
considered scalar-type perturbations with $k^2 = 0$. 
It is not immediately obvious that this would give equivalent
results, since when $k^2 = 0$, the quantities $\scalar_{i}$ and $\scalar_{ij}$ 
do not exist, so $B$ and $H_T$ are not defined and the $x^i$ coordinate freedom
in eq.~(\ref{coord}) does not exist. Furthermore,
eqs.~(\ref{Ein:ij:tracefree}) become trivial, and therefore 
the relation, eq.~(\ref{rel:Psi:Phi:X}) need not hold. Thus, it is not
entirely straightforward to make a physical
correspondence between perturbations with
$k^2=0$ and the $k^2 \rightarrow 0$ limit of perturbations with $k^2 \neq 0$.
Nevertheless, such a one-to-one, onto correspondence does exist\footnote{ 
It also is worth pointing out that, 
although the gauge freedom is different, the effective stress-energy
tensor for $k^2=0$ perturbations remains highly gauge dependent.
 }, 
and can be explicitly achieved
by using the gauge freedom available when $k^2 \neq 0$ to set $B = H_T =0$
and using the gauge freedom available when $k^2 = 0$ to set $A = - H_L$.
Further discussion of the relationship between perturbations 
in the long-wavelength limit and exactly homogeneous perturbations can 
be found in Refs.~\cite{NT1998,KH1998,ST1998}. 

Since scalar-type perturbations with $k^2=0$ manifestly correspond 
to perturbations to other FLRW spacetimes, it is clear that one cannot 
find any new physical phenomena that are not already present in FLRW models 
by studying long-wavelength perturbations and dropping all terms 
that are $O(k^2)$. 
For example, consider a FLRW model which contains two matter components, 
such as dust and radiation or two scalar fields. 
In such a model, there exist nontrivial gauge-invariant perturbations 
even in the $k\rightarrow 0$ limit, and implications of such perturbations 
to the cosmological back-reaction problem have been discussed 
in~\cite{Nambu05a,AW01,GB05}.  
However, in the $k\rightarrow 0$ limit such perturbations merely correspond 
to perturbations to other FLRW models; in the above examples, they would 
correspond to changing the proportion of dust and radiation or changing 
the initial conditions of the scalar fields. These perturbations cannot 
give rise to any new phenomena---such as physically measurable 
acceleration---that are not already present in exact FLRW models. 

\section{Summary}  
In this paper, we have argued that the attempts to explain cosmic
acceleration by effects of inhomogeneities, without invoking a
cosmological constant or dark energy, are, at best, highly
implausible. A Newtonianly perturbed FLRW metric appears to describe
our universe very accurately on all scales. In this model, the
back-reaction effects of inhomogeneities on the cosmological dynamics
are negligible even though the density contrast may be very large on
small scales.

We focused much of our attention on exposing the flaws in 
two types of attempts to explain acceleration by effects of inhomogeneities. 
(i) Starting from an inhomogeneous model, 
one can obtain an effective FLRW universe by spatial averaging. This effective
FLRW universe may display acceleration. However, we showed explicitly via
concrete examples that acceleration of the effective FLRW universe may 
occur in situations
where no physically observable effects of acceleration actually occur.
(ii) The back-reaction effects of a perturbation of a FLRW universe
are described at second order by an effective stress-energy tensor 
constructed from the first order perturbation. In particular cases, this
effective stress-energy tensor may take 
the form of a cosmological constant, thereby suggesting that it could produce
acceleration. However, we pointed out that
(unlike the effective stress-energy tensor arising in the shortwave 
approximation), the effective stress-energy tensor arising in second order
perturbation theory is highly gauge dependent and must be small in order
to justify neglecting higher order corrections. We explicitly evaluated the 
second-order effective stress-energy tensor for pure gauge 
scalar-type perturbations of an Einstein-scalar field model, 
and showed that it can take essentially any form one wishes, 
including the form of a cosmological constant. 

\bigskip 

\begin{center} 
{\bf Acknowledgments} 
\end{center} 

This research was supported by NSF grant PHY 00-90138 to the University 
of Chicago. 
\bigskip 

\appendix 

\section*{Appendix} 
Here we provide some details of the construction of the accelerating
domain $\D$ in Minkowski spacetime that was described below
eq.~(\ref{average:acceleration}).  Let $t$ and $r$ be, respectively,
the standard time and radial coordinates in Minkowski spacetime.  Let
$a>0$ and $r_0>\epsilon>0$.  Let $f$ be a smooth, monotone decreasing
function of one variable such that $f(x) = 1$ for all $x \leq 1/2$,
$f(x) = 0$ for all $x \geq 1$. Define 
\ben
\psi(r) = f \left(\frac{r-r_0}{2\epsilon} +1 \right) \,. 
\een
Then $\psi(r) = 1$ whenever $r\leq r_0-\epsilon$ and $\psi(r) = 0$ 
whenever $r \geq r_0$. Furthermore, 
there exists 
a constant $C>0$, independent of $\epsilon$, such that $\epsilon |\psi'| < C $ 
for all $0<\epsilon<r_0$, where $\psi' \equiv \partial \psi/\partial r$. 
Define
\ben 
F(r) \equiv \left\{\psi(r) - \psi(-r + 2r_0 ) \right\} 
             \left( \sqrt{r^2 + a^2} - \sqrt{r_0^2 + a^2} \right) 
             + \sqrt{r_0^2 + a^2} \,.     
\label{embedding}
\een 
Then $F$ is smooth and the hypersurface $\Sigma$ defined by $t=F(r)$ also
is smooth. For $r < r_0 - \epsilon$, $\Sigma$ is the expanding hyperboloid
$t = \sqrt{r^2 + a^2}$, whereas for $r > r_0 + \epsilon$, $\Sigma$ 
is the contracting hyperboloid $t = - \sqrt{r^2 + a^2} + 2\sqrt{r_0^2 + a^2}$.
The local expansion rate $H\equiv \dot a/a$ of 
$\Sigma$ smoothly changes from $1/a$ to $-1/a$ in the junction interval 
$(r_0 -\epsilon , \,r_0+\epsilon)$. 
Nowhere does $\Sigma$ display an accelerated expansion locally. 

Now let us take our domain $\D$ to be a ball of radius $r_\D$ on $\Sigma$,
where $r_\D$ is chosen so that $\dot a_\D$ vanishes. 
It is always possible to find such an $r_\D$ since, 
by construction, $V_\D$, hence $\dot a_\D $, is a smooth function 
of $r$, and the local expansion rate $H=\dot a/a$ is positive when 
$r< r_0-\epsilon$, whereas it is negative when $r_0+\epsilon < r$. 
Since the third term of eq.~(\ref{average:acceleration}) vanishes 
for Minkowski spacetime, if $\langle {\cal R} \rangle_\D$ is negative in $\D$, 
then eq.~(\ref{average:acceleration}) shows an acceleration 
$3\ddot a_\D/a_\D = -\langle {\cal R}\rangle_\D >0$. 
However, apart from the junction 
region $(r_0 -\epsilon, \, r_0 + \epsilon)$, 
$\Sigma$ is intrinsically a hyperbolic space with a negative scalar 
curvature ${\cal R} = -6/a^2$. Furthermore, the following 
calculation shows that the contribution to 
$\langle {\cal R} \rangle_\D$ from the junction region can be 
made negligibly small. The induced metric on $\Sigma$ is 
\ben
 ds^2 = \left( 1-F'^2 \right)dr^2 + r^2 d\Omega^2 \,, 
\een 
so the scalar curvature of the junction region is given by
\ben
 {\cal R} = -\frac{4r F'F'' + 2(F')^2 \left\{1- (F')^2 \right\} 
                  }{r^2 \left\{1- (F')^2  \right\}^2 }\,. 
\een 
Using the formula, 
\ben
 F' = - \frac{r_0}{\sqrt{r_0^2 + a^2}} (\psi + \epsilon \psi') 
            + O(\epsilon) \,,  
\een
which is obtained from the properties of $\psi$ in the junction interval,  
one finds 
\bena
 \int_{r_0-\epsilon}^{r_0+\epsilon} dr r^2 \sqrt{1- (F')^2} {\cal R} 
 &=&  \int_{r_0-\epsilon}^{r_0+\epsilon} dr 
       \left\{ 
              2\frac{2-(F')^2}{\sqrt{1- (F')^2}} 
              - 4\frac{\partial }{\partial r}
                 \left( \frac{r}{\sqrt{1- (F')^2}}\right) 
       \right\} 
\non \\
 &=& O(\epsilon) \,,     
\eena 
which can be made arbitrarily small by taking $\epsilon \rightarrow 0$.  
Thus, the contribution to $\langle {\cal R} \rangle_\D$
from the junction region can indeed be made arbitrarily small,
so that $-\langle {\cal R}\rangle_\D \approx 6/a^2 > 0$.  
Thus, one obtains an accelerating domain $\D \subset \Sigma$ 
in Minkowski spacetime through 
the volume averaging process. 



\begin{thebibliography}{99} 

\bibitem{Buchert00}
Buchert,~T., 2000, Gen.~Rel.~Grav. {\bf 32}, 105.  

\bibitem{Buchert01}  
Buchert,~T., 2001, Gen.~Rel.~Grav. {\bf 33}, 1381.  

\bibitem{BC03} 
Buchert,~T. and  Carfora,~M., 2003 Phys.~Rev.~Lett. {\bf 90}, 031101.  

\bibitem{Rasanen03} 
Rasanen,~S., 2004, JCAP 0402 003.  

\bibitem{KMNR05}
Kolb, E.W., Matarrese, S., Notari, A., and Riotto, A.,  
hep-th/0503117.  

\bibitem{KMR05} 
Kolb, E.W., Matarrese, S., and Riotto, A.,  
astro-ph/0506534.  

\bibitem{BMR05}
Barausse, E., Matarrese, S., and Riotto, A., 2005 
Phys. Rev. D {\bf 71}, 063537.  

\bibitem{Nambu05a}
Nambu,~Y., 2005 Phys. Rev. D {\bf 71}, 084016.  

\bibitem{NT05} 
Nambu,~Y. and Tanimoto,~M., gr-qc/0507057.  

\bibitem{Moffat05}
Moffat,~J.W., astro-ph/0505326. 

\bibitem{MAB1997Lett}
Mukhanov, V.F. Abramo, L.R.W., and Brandenberger, R.H., 1997
Phys. Rev. Lett. {\bf 78}, 1624.  

\bibitem{ABM1997PRD}
Abramo, L.R.W., Brandenberger, R.H., and
Mukhanov, V.F., 1997 Phys. Rev. D {\bf 56}, 3246.  


\bibitem{Nambu02}
Nambu,~Y., 2002 Phys. Rev. D {\bf 65}, 104013.  

\bibitem{BL2004} 
Brandenberger, R.H. and Lam, C.S., hep-th/0407048.  


\bibitem{Unruh1998}
Unruh, W., 1998 astro-ph/9802323.  

\bibitem{BH1964} 
Brill, D. and Hartle, J., 1964 Phys. Rev. {\bf 135}, B271. 

\bibitem{Isaacson1968} 
Isaacson, R., 1968 Phys. Rev. {\bf 166}, 1272.  

\bibitem{Flanagan05} 
Flanagan, E.E., 2005 Phys. Rev. D {\bf 71}, 103521.  

\bibitem{HS05}
Hirata, C.M. and Seljak, U., 2005 astro-ph/0503582. 

\bibitem{GCA05} 
Geshnizjani, G., Chung, D.J.H., and Afshordi, N., 
2005 Phys. Rev. D {\bf 72}, 023517. 


\bibitem{w84} Wald,~R.M., 1984 {\it General Relativity}, 
University of Chicago Press: Chicago. 

\bibitem{FS83}
Futamase,~T. and Schutz,~B.F., 1983 
Phys. Rev. D {\bf 28}, 2363. 
 
\bibitem{ehlers}
Ehlers,~J., 1997 Class. Quant. Grav. {\bf 14}, A119. 

\bibitem{HolzWald97} 
Holz, D.E. and Wald, R.M., 1998 Phys. Rev. D {\bf 58}, 063501.  

\bibitem{Barrow88}
Barrow,~J.D., 1988 
Quart. J. Roy. astr. Soc., {\bf 30}, 163. 

\bibitem{Futamase89}
Futamase,~T., 1989 Mon. Not. R. astr. Soc. {\bf 237}, 187. 

\bibitem{Zalaletdinov97}
Zalaletdinov,~R.M., 1997 Bull.~Astron.~Soc.~India {\bf 25}, 401. 

\bibitem{CPZ05}
Coley,~A.A., Pelavas,~N. and Zalaletdinov,~R.M., gr-qc/0504115. 

\bibitem{Buchert05}
Buchert,~T., gr-qc/0507028.  

\bibitem{MTW} 
Misner, C. W., Thorne, K. S. and Wheeler, J.A., 1973 
{\it Gravitation} (Freeman, San Francisco).  

\bibitem{Burnett1989}
Burnett, G.A., 1989 J. Math. Phys. {\bf 30}, 90. 

\bibitem{AW01}
Abramo,~L.R. and Woodard,~R.P., 
2002 Phys. Rev. D {\bf 65}, 043507.

\bibitem{GB05}
Geshnizjani,~G. and Brandenberger, R.H., 
2005 JCAP {\bf 04}, 006.  

\bibitem{Bardeen1980}
Bardeen, J.M., 1980 Phys. Rev. D {\bf 22}, 1882.

\bibitem{KS1984}
Kodama,~H. and Sasaki,~M., 1984 Prog. Theor. Phys. Supple. {\bf 78}, 1. 

\bibitem{NT1998} 
Nambu,~Y. and Taruya,~A,~1998 Class. Quant. Grav. {\bf 15}, 2761.  

\bibitem{KH1998} 
Kodama, H. and Hamazaki, T., 1998 Phys. Rev. D {\bf 57}, 7177. 

\bibitem{ST1998}
Sasaki,~M. and Tanaka,~T., 1998 Prog. Theor. Phys. {\bf 99}, 763. 

\end{thebibliography}
\end{document}